\documentclass[twocolumn,prl]{revtex4}
\usepackage{float}
\usepackage{makeidx}
\usepackage{graphicx,amsmath}
\usepackage{CJK}

\begin{document}
\begin{CJK*}{GB}{gbsn}

\title{Bose-Einstein Condensate in a light-induced vector gauge potential using the 1064 $nm$ optical dipole trap lasers}
\author{Zhengkun Fu, Pengjun Wang, Shijie Chai, Lianghui Huang, Jing Zhang$^{\dagger}$}
\affiliation{State Key Laboratory of Quantum Optics and Quantum
Optics Devices, Institute of Opto-Electronics, Shanxi University,
Taiyuan 030006, P.R.China \label{in}}

\begin{abstract}

Using two crossed 1064 $nm$ optical dipole trap lasers to be the
Raman beams, an effective vector gauge potential for Bose-Einstein
condensed $^{87}$Rb in the $F=2$ hyperfine ground state is
experimentally created. The moderate strength of the Raman coupling
still can be achieved when the detuning from atomic resonance is
larger than the excited-state fine structure, since rubidium has 15
nm energy-level spitting. The atoms at the far detuning of the Raman
coupling are loaded adiabatically into the dressed states by ramping
the homogeneous bias magnetic field with different pathes and the
dressed states with different energies are studied experimentally.
The experimental scheme can be easily extended to produce the
synthetic magnetic or electric field by means of a spatial or time
dependence of the effective vector potential.

\end{abstract}

\maketitle
\end{CJK*}

Quantum degenerate gases in ultracold temperature offer us new
opportunities to efficiently simulate quantum condensed matter
systems \cite{one}. It is an important advantage for experiments
that the physical parameters of atomic systems, including the number
of the trapped atoms, the shape of the trapping potential, and the
strength of the atom-atom interaction can be precisely controlled. A
fascinating example of utilizing atomic systems is that the vector
potential of the charged particles in a magnetic field can be
simulated by the ultracold atomic gas if a gauge field is applied on
it. In this case, the strongly correlated states of matter, such as
the fractional quantum Hall effect exhibited by electrons in a
magnetic field, can be easily studied. A well-known method is to
rotate the gas \cite{Fetter,Cooper}, where the transformation to the
rotating frame corresponds to giving the particles a fictitious
charge, and applying an effective uniform magnetic field. Another
approach is to induce gauge potentials through the laser-atom
interactions \cite{three,Dalibard}. There are already various
theoretical proposals for generating artificial abelian and
non-abelian gauge fields without \cite{four,five,six,seven,eight} or
with optical lattices
\cite{nine,ten,eleven,twelve,thirteen,forteen,fifteen}, and some
exotic properties are predicted
\cite{four,five,six,seven,eight,nine,ten,eleven,twelve,thirteen,forteen,fifteen,sixteen,seventeen,eighteen,ninteen}.
The experiment on the generation of synthetic gauge fields has had
made great progress recently in the NIST group
\cite{twenty,twenty-one,twenty-two,twenty-three}. In Lin et al.
experiment \cite{twenty}, the effective vector potential is
generated by coupling a pair of 804.3 $nm$ Raman laser beams into
the magnetic sublevels of the $F=1$ hyperfine level of the
electronic ground state in a 1550 $nm$ optically trapped
Bose-Einstein condensate (BEC) of $^{87}$Rb atoms. Successively, the
synthetic magnetic \cite{twenty-one} and electric fields
\cite{twenty-two} were also produced from a spatial variation and
time dependence of the effective vector potential. Very recently,
using the similar scheme BEC with spin-orbit coupling
\cite{twenty-three} has also been realized by the same group.

In this letter, we report a novel experimental scheme of generating
light-induce vector gauge potential, in which the two 1064 $nm$
optical dipole trap lasers are also used as a pair of Raman lasers
in $^{87}$Rb BEC. At first an optically trapped BEC is created by
the two crossed optical dipole trap lasers. Simultaneously, the two
dipole trap lasers with a frequency difference resonant with the
energy difference between the magnetic sublevels, couple these two
magnetic sublevels of the $F=2$ hyperfine level of the electronic
ground state. We adiabatically load the atoms at the far detuning of
the Raman coupling into the dressed state by ramping the bias
magnetic field to resonance. The different energy dressed states are
loaded and studied. The collision decay of the high energy dressed
state is observed. The light-induce vector gauge potential by
four-photon Raman process with $4\hbar k_{R}$ momentum is also
observed. Our experimental setup can be easily extended to present
the spatial or time dependent vector potential for producing
synthetic magnetic or electric field.

\begin{figure}
\centerline{
\includegraphics[width=3in]{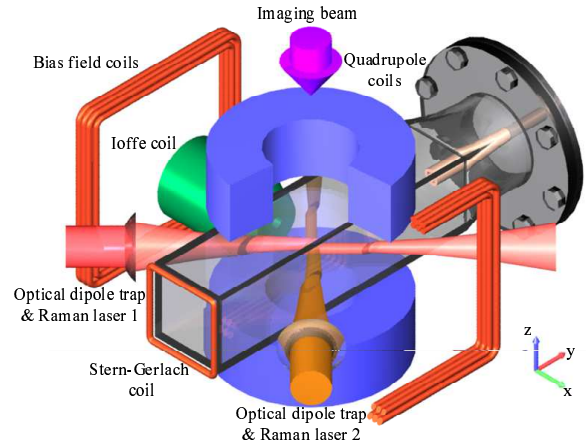}
} \vspace{0.1in}
%\setlength{\columnwidth}{3.2in}
%\centerline{
\caption{(Color online). Schematic drawing of the experimental
setup. \label{Fig1} }
%}
\end{figure}

The model with two-level system in Ref. \cite{eight,twenty-three} is
employed in the experiment \cite{supp}. We choose the two magnetic
sublevels $|\uparrow\rangle=|F=2,m_{F}=2\rangle$ and
$|\downarrow\rangle=|F=2,m_{F}=1\rangle$ of the $F=2$ hyperfine
level of the electronic ground state to be the two internal spin
states, which are coupled by a pair of Raman beams with strength
$\Omega$. In our experiment, the optical dipole trap is composed of
two horizontal crossed beams at $90^{o}$ along $\hat{x}\pm \hat{y}$
and overlapped at the focus, which also are used as two Raman beams,
as shown in Fig. 1. The linear polarization of both beams is
horizontal in the plane of x-y. Both beams are extracted from a 15
$W$ laser (MOPA 15 NE, InnoLight Technology, Ltd.) operating at the
wavelength of 1064 $nm$ with the narrow linewidth single-frequency.
Two beams single-pass through two acousto-optic modulators (AOM)
(3110-197, Crystal Technology, Inc.). The Raman beam 1 and 2 are
frequency shifted -100 $MHz$ and -110.4 $MHz$ by two signal
generators (N9310A, Agilent) respectively. The frequency difference
$\nu_{R}/2\pi=10.4$ $MHz$ of two signal generators are phase-locked
by a source locking CW microwave frequency counters (EIP 575B, Phase
Matrix Inc.). Thus two laser beams are phase-locked and frequency
shifted 10.4 $MHz$ relative to each other to avoid any spatial
interference between the two beams, and at the same time provide the
radio-frequency Raman coupling between two magnetic sublevels. Then
two beams are coupled into two high power polarization maintaining
single-mode fibers in order to increase stability of the beam
pointing and obtain better beam-profile quality. Behind the fibers,
one beam is focused to a waist size of $1/e^{2}$ radii of 38 $\mu m$
by a achromatic lens of focus length $f=300$ $mm$, and the other
beam is focused to 49 $\mu m$ by a $f=400$ $mm$ lens. For enhancing
the intensity stability of the two beams, a small fraction of the
light is sent into a photodiode and the regulator is used for
comparing the intensity measured by the photodiode to a set voltage
value from the computer. The non-zero error signal is compensated by
adjusting the radio power in the AOM in front of the optical fiber.

A homogeneous bias magnetic field provided by a pair of Helmholtz
coils along $\hat{y}$ gives a linear Zeeman shift $\omega_{Z}/2\pi$
between two magnetic sublevels, as shown in Fig. 1. To control the
magnetic field precisely and reduce the magnetic field noise, the
power supply (Delta SM70-45D) has been operated in remote voltage
programming mode, whose voltage is set by an analog output of the
experiment control system. The current through the coils is
controlled by the external regulator relying on a precision current
transducer (Danfysik ultastable 867-60I). The output error signal
from the regulator actively stabilize the current with the PID (
proportional-integral-derivative) controller acting on the MOSFET
(metal-oxide-semiconductor field-effect transistor). In order to
reduce the current noise and decouple the control circuit from the
main current, a conventional battery is used to power the circuit.

In our experiment, $^{87}$Rb atoms are first precooled to 1.5 $\mu
K$ by radio-frequency evaporative cooling in a quadrupole-Ioffe
configuration (QUIC) trap \cite{twenty-four,twenty-five}.
Subsequently, the atom sample is transferred back to the center of
the glass cell \cite{twenty-six} in favor of the optical access.
After loading  $^{87}$Rb atoms in hyperfine state
$|F=2,m_{F}=2\rangle$ into the dipole trap with the full powers (900
$mW$ and 1.3 $W$) at a weak homogeneous bias magnetic field about 1
$G$, the forced evaporative cooling is performed by lowering the
powers of two beams \cite{twenty-seven}. With the evaporation time
of 1.2 $s$, the pure BEC with the atomic number of $2\times10^{5}$
is obtained at the powers of 169 $mW$ (beam 1) and 320 $mW$ (beam
2). In order to increase the Raman coupling strength, the powers of
two dipole trap beams are increased to 207 $mW$ (beam 1) and 480
$mW$ (beam 2), respectively. The pure BEC is still maintained in the
dipole trap with trap frequency of $2\pi\times83$ $Hz$ along
$\hat{x}$ and $\hat{y}$. Now we first measure the resonant Raman
Rabi frequency $\Omega$ by observing population oscillations driven
by the variable Raman pulse length. The third dipole trap beam (beam
3) with frequency shifted -103 $MHz$ counterpropagating with the
Raman beam 2 is utilized in the measurement process. The BEC is
loaded adiabatically into the crossed dipole trap composed of Raman
beam 1 and dipole trap beam 3 by ramping the dipole trap beam 3 and
decreasing the intensity of the Raman beam 2 to zero. Then the
homogeneous bias magnetic field is ramped to the value with
$\hbar\delta=-4E_{R}$, so the atoms are resonant for
$|\uparrow',0\rangle \rightarrow |\downarrow',-2k_{R}\rangle$ (the
energy gap $E_{+}(p=-k_{R})-E_{-}(p=-k_{R})=\hbar\Omega$). By the
variable Raman pulse length of Raman beam 2, the observed
oscillation period of 420 $\mu s$ corresponds to the resonant Raman
Rabi frequency $\hbar\Omega=2.35E_{R}$.

\begin{figure}
\centerline{
\includegraphics[width=2.5in]{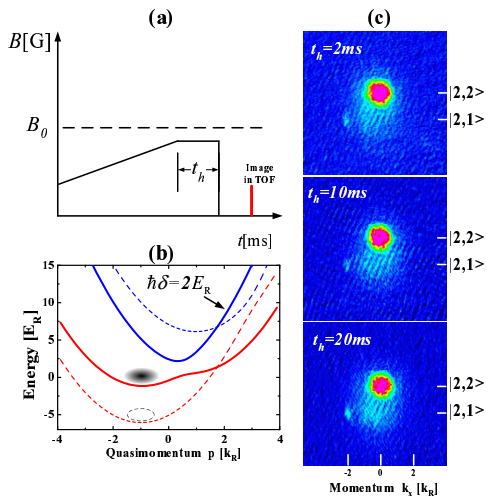}
} \vspace{0.1in}
%\setlength{\columnwidth}{3.2in}
%\centerline{
\caption{(Color online). (a) The time sequence of the homogeneous
bias magnetic field for loading the atoms into the low-energy
Raman-dressed state. The horizontal dashed line indicates $B_{0}$,
which corresponds to $\delta=0$. (b) Energy-quasimomentum dispersion
for $\hbar\delta=2E_{R}$ (thick solid curves) and $\hbar\delta\gg
E_{R}$ (thin dashed curves) at Raman coupling strength
$\hbar\Omega=2.35E_{R}$. The quasimomentum of BEC keeps in the
low-energy dressed state with $p_{min}<0$ for $B<B_{0}$. (c) Images
(1.17 $mm$ by 1.17 $mm$) of the Raman-dressed state for
$\hbar\delta=2E_{R}$ with variable hold times $t_{h}$ after a 30
$ms$ TOF. The two spin and momentum components,
$|\uparrow,k_{x}=p_{min}+k_{R}\rangle$ and
$|\downarrow,k_{x}=p_{min}-k_{R}\rangle$ \cite{supp}, are separated
along $\hat{y}$ (a Stern-Gerlach field is applied along $\hat{y}$
before the image). \label{Fig2} }
%}
\end{figure}

We adiabatically load the BEC initially in $|\uparrow,0\rangle$ into
the Raman-dressed states of the low $E_{-}$ or high energy $E_{+}$,
simply by ramping the homogeneous bias magnetic field with the
different pathes. Here, when the atoms are Raman resonant (at 10.4
$MHz$ with $\delta=0$) between $|\uparrow\rangle$ and
$|\downarrow\rangle$, the detuning between $|\downarrow\rangle$ and
$|F=2,m_{F}=0\rangle$ is about $-30E_{R}$, we may regard it as the
two-level system. At last, the Raman dressed states may be
characterized by the time-of-flight (TOF). When the Raman beams and
the homogeneous bias magnetic field are turned off abruptly, the
atoms are projected onto its individual spin and momentum
components. The atoms then expand in a magnetic field gradient for
28 $ms$ during TOF along $\hat{y}$, and the two spin states are
separated spatially due to the Stern-Gerlach effect. Imaging the
atoms after a 30 $ms$ TOF gives the momentum and spin composition of
the dressed state. Now we discuss three cases of loading the BEC
into the Raman-dressed states by ramping the homogeneous bias
magnetic field with three different paths.

\begin{figure}
\centerline{
\includegraphics[width=2.5in]{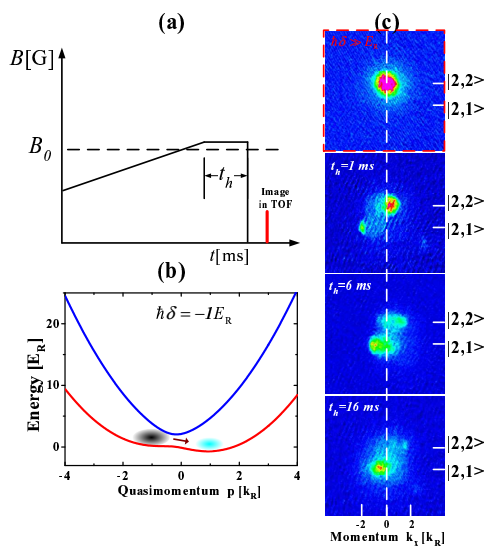}
} \vspace{0.1in}
%\setlength{\columnwidth}{3in}
%\centerline{
\caption{(Color online). (a) The time sequence for loading the atoms
into the low-energy Raman-dressed state. (b) Energy-quasimomentum
dispersion for $\hbar\delta=-E_{R}$ (thick solid curves). The
dressed atoms locating at high-energy well of the double wells are
unstable and will transit to the low-energy well. (c) Images of the
Raman-dressed state for $\hbar\delta=-E_{R}$ with variable hold
times $t_{h}$ after a 30 $ms$ TOF. \label{Fig3} }
%}
\end{figure}

$Case$ $1$: We prepare the BEC initially in $|\uparrow,0\rangle$
locating in the low energy branch $E_{-}$ with the far positive
detuning $\hbar\delta \gg E_{R}$ by setting the homogeneous bias
magnetic field at the value of $B\ll B_{0}$, as shown in Fig. 2(a)
and (b). Here, $B_{0}$ corresponds to the $\delta=0$. Then we ramp
the homogeneous bias magnetic field slowly in a time 150 $ms$ to the
value with $\hbar\delta=2E_{R}$ and hold on in a variable time
$t_{h}$. Since $\Omega<4E_{R}$ in the experiment, the low energy
$E_{-}(p)$ presents the double wells in quasi-momentum space. When
$\hbar\delta=2E_{R}$, the double wells become asymmetry and the
low-energy well locates at $p_{min}=-0.925k_{R}$. Thus the atoms are
loaded to low-energy dressed state adiabatically and locate
low-energy well of the double wells at $p_{min}=-0.925k_{R}$. Figure
2(c) shows spin-resolved TOF images of adiabatically loaded the
dressed state with the different holding times. These images
demonstrate that the atoms are loaded to low-energy dressed state
adiabatically at the low-energy well of the double wells, which are
very stable with the long-life time.

$Case$ $2$: The initial condition is same as the case 1. The
difference is that the homogeneous bias magnetic field is ramped to
the value with $\hbar\delta=-E_{R}$ ($B>B_{0}$) as shown in Fig.
3(a). The low-energy well of the asymmetry double wells is changed
into $p_{min}=0.889k_{R}$. The atoms still are loaded to low-energy
dressed state adiabatically, however locate at high-energy well (no
$p_{min}$) of the double wells as shown Fig. 3(b). The dressed atoms
locating at high-energy well of the double wells are unstable and
will transit to the low-energy well. Images in Fig. 3(c) show this
transition process. After holding time of 20 $ms$, the dressed atoms
populate the low-energy well of the double wells.

\begin{figure}
\centerline{
\includegraphics[width=2.5in]{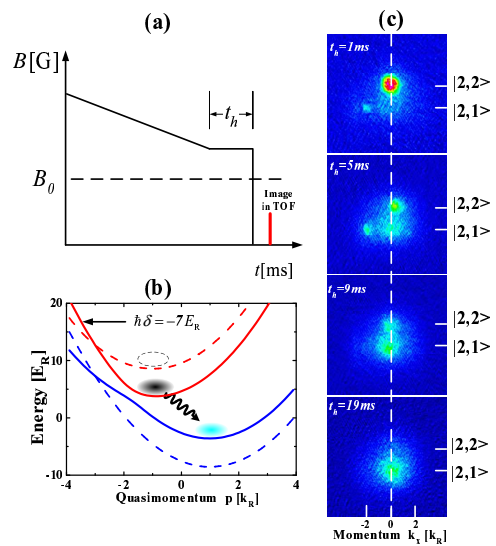}
} \vspace{0.1in}
%\setlength{\columnwidth}{3in}
%\centerline{
\caption{(Color online). (a) The time sequence for loading the atoms
into the high-energy Raman-dressed state. (b) Energy-quasimomentum
dispersion for $\hbar\delta=-7E_{R}$ (thick solid curves) and
$\hbar\delta\ll-7E_{R}$ (thin dashed curves). The atoms start in
high-energy Raman-dressed state and ultimately decay into the
low-energy Raman-dressed state. (c) Images of the Raman-dressed
state for $\hbar\delta=-7E_{R}$ with variable hold times $t_{h}$
after a 30 $ms$ TOF. \label{Fig4} }
%}
\end{figure}

$Case$ $3$: We prepare the BEC initially in $|\uparrow,0\rangle$
locating in the high energy branch $E_{+}$ with the far negative
detuning $\hbar\delta \ll -E_{R}$ by setting the homogeneous bias
magnetic field at the value of $B>B_{0}$, as shown in Fig. 4(a) and
(b). The homogeneous bias magnetic field is decreased to the value
with $\hbar\delta=-7E_{R}$ and the atoms are loaded to high-energy
dressed state adiabatically. The high energy branch $E_{+}(p)$
consists of single well in quasi-momentum space, thus the dressed
atoms locate at $p_{min}=-0.84k_{R}$. The dressed atoms in high
quasibands are energetically allowed, however, collisional decay
will present near Raman resonance except the lowest-energy dressed
state \cite{eight}. The decay for variable hold times ranging from 1
$ms$ to 19 $ms$ is observed as shown in Fig. 4(c). The dressed atoms
in high quasibands decay into the low-energy band accompanying the
heating, which is a completely different process compared with that
of case 2.

\begin{figure}
\centerline{
\includegraphics[width=2.5in]{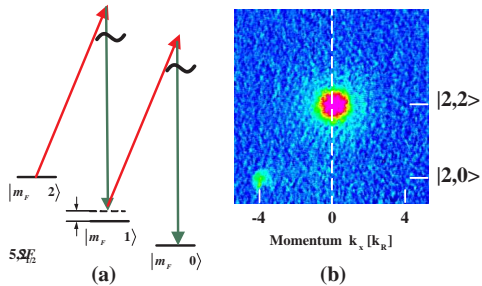}
} \vspace{0.1in}
%\setlength{\columnwidth}{3in}
%\centerline{
\caption{(Color online). (a) The scheme for generating four-photon
Raman process. (b) Images of the four-photon Raman-dressed state
after a 30 $ms$ TOF. The two spin and momentum components,
$|\uparrow,k_{x}=p_{min}+2k_{R}\rangle$ and
$|\tilde{\downarrow},k_{x}=p_{min}-2k_{R}\rangle$, are separated
along $\hat{y}$. \label{Fig5} }
%}
\end{figure}

Moreover, the four-photon Raman process with $4\hbar k_{R}$ momentum
is observed, which may be used to generate the light-induce vector
gauge potential consisting of the high momentum components. The
similar method has been done \cite{twenty-eight} that the cold atoms
were coherently transferred from one dressed state to another one by
a multi-photon process, which changed the atom momentum by several
photon recoils. When the atoms are Raman red detuning with
$\delta=-15E_{R}$ between $|\uparrow\rangle$ and
$|\downarrow\rangle$, so the blue detuning between
$|\downarrow\rangle$ and $|F=2,m_{F}=0\rangle$ is $+15E_{R}$, the
condition for the four-photon resonant Raman process
($2\nu_{R}=\omega_{Z}^{|2,2\rangle\leftrightarrow|2,0\rangle}$) is
satisfied as shown in Fig. 5(a). The nonzero two-photon detuning
$\delta=\pm15E_{R}$ is used to suppress resonant two-photon Raman
process. Therefore, the spin state $|F=2,m_{F}=1\rangle$ has
negligible contribution in case of the large detuning. We may regard
it as the two-level system consisting of $|F=2,m_{F}=2\rangle$ and
$|F=2,m_{F}=0\rangle=|\tilde{\downarrow}\rangle$. When ramping the
homogeneous bias magnetic field slowly from low field to the
four-photon Raman resonance, the atoms are loaded to low-energy
dressed state. The two spin and momentum components (
$|\uparrow,k_{x}=p_{min}+2k_{R}\rangle$ and
$|\tilde{\downarrow},k_{x}=p_{min}-2k_{R}\rangle$) for the dressed
state are observed as shown in Fig. 5(b). It will be useful to
produce the large size of double wells in quasi-momentum space.

In conclusion, we have demonstrated an effective vector gauge
potential for $^{87}$Rb BEC in the $F=2$ hyperfine ground state,
which was generated by using two crossed 1064 $nm$ optical dipole
trap lasers to be the Raman beams. The effective vector gauge
potential still can be generated (in the atomic long lifetime due to
photon scattering in the optical dipole trap) by the very
far-detuning (larger than the excited-state fine structure spitting)
between the single-photon resonance and the excited state
transition, since rubidium has 15 nm the excited-state fine
structure spitting \cite{eight,supp}. The experimental scheme can be
applied to produce the synthetic magnetic or electric field by means
of a spatial or time dependence of the effective vector potential.

$^{\dagger}$Corresponding author email: jzhang74@yahoo.com,
jzhang74@sxu.edu.cn

\begin{acknowledgments}
This research is supported by National Basic Research Program of
China (Grant No. 2011CB921601), and NSFC (Grant No. 10725416,
60821004).
\end{acknowledgments}

\title{SUPPLEMENTARY MATERIAL}

\maketitle

$Two-level$ $model$: We describe in detail theoretical model of
two-level system for generating light-induce vector gauge potential.
Two Raman beams have frequencies $\omega_{R}$ and
$\omega_{R}+\nu_{R}$, and a bias field $B$ along $\hat{x}$ produces
a Zeeman shift $\hbar\omega_{Z}=g\mu_{B}B$. Since the momentum
transfer induced by the Raman beams is along $\hat{x}$, the
Hamiltonian is written as
$H=H_{R}(k_{x})+[\hbar^{2}(k_{y}^{2}+k_{z}^{2})/2m+V(\bar{r})]$,
where $H_{R}(k_{x})$ is the Hamiltonian for the Raman coupling, the
Zeeman energies and the motion along $\hat{x}$. $V(\bar{r})$ is the
state-independent trapping potential arising from the scalar light
trap of the Raman beams and $m$ is the atomic mass. Under the
rotating-wave approximation in the frame rotating at $\nu_{R}$, the
Hamiltonian $H_{R}(k_{x})$ is written in the bare atomic state basis
of
$\{|\uparrow,k_{x}=p+k_{R}\rangle,|\downarrow,k_{x}=p-k_{R}\rangle\}$

\begin{eqnarray}
H_{R}(k_{x})=\hbar \left(
  \begin{array}{cc}
     \frac{\hbar}{2m}(p+k_{R})^{2}-\delta/2 & \Omega/2 \\
    \Omega/2 &  \frac{\hbar}{2m}(p-k_{R})^{2}+\delta/2 \\
  \end{array}
\right).
\end{eqnarray}
Here $\delta=\nu_{R}-\omega_{Z}$ is the detuning from Raman
resonance, $\Omega$ is the resonant Raman Rabi frequency, $p$
denotes quasimomentum. $k_{R}=k_{r}\sin(\theta/2)$, $k_{r}=2\pi
/\lambda$ is the single-photon recoil momentum, $\lambda$ is the
wavelength of the Raman beam, and $\theta=90^{o}$ is the
intersecting angle of two Raman beams. $\hbar k_{R}$ and
$E_{R}=(\hbar k_{R})^{2}/2m = h\times 1.013$ $kHz$ are the units of
momentum and energy, respectively. $H_{R}(k_{x})$ are diagonalized
to get two energy eigenvalues
$E_{\pm}(p)=\hbar[\hbar(p^{2}+k_{R}^{2})/2m \pm \sqrt{(4\hbar p
k_{R}/2m -\delta)^{2}+\Omega^{2}}/2]$, which give the effective
dispersion relations of the dressed states. The two dressed
eigenstates are expressed by
\begin{eqnarray}
|\uparrow',p\rangle&=&c_{1}|\uparrow,k_{x}=p+k_{R}\rangle+c_{2}|\downarrow,k_{x}=p-k_{R}\rangle \nonumber \\
|\downarrow',p\rangle&=&c_{3}|\uparrow,k_{x}=p+k_{R}\rangle+c_{4}|\downarrow,k_{x}=p-k_{R}\rangle.
\end{eqnarray}
Here, $c_{1}=1/\sqrt{a^2+1}$, $c_{2}=a/\sqrt{a^2+1}$, and
$a=-(4\hbar p k_{R}/2m -\delta-\sqrt{(4\hbar p k_{R}/2m
-\delta)^{2}+\Omega^{2}})/\Omega$. $c_{3}=1/\sqrt{b^2+1}$,
$c_{4}=b/\sqrt{b^2+1}$, and $b=-(4\hbar p k_{R}/2m
-\delta+\sqrt{(4\hbar p k_{R}/2m -\delta)^{2}+\Omega^{2}})/\Omega$.
$|\uparrow',p\rangle$ is the high-energy dressed state for
$E_{+}(p)$ and $|\downarrow',p\rangle$ is the low-energy dressed
state for $E_{-}(p)$. Since the high and low energies $E_{\pm}$ of
the dressed states depend on the experimental parameters $\Omega$
and $\delta$, the positions of energy minima ($p_{min}$) are thus
experimentally tunable. For $\Omega<4E_{R}$ and small $\delta$, the
lowest energy $E_{-}(p)$ consists of double wells in quasi-momentum
space. As $\Omega>4E_{R}$, the double wells merge into a single
well.

$Raman$ $coupling$ $strength$ and $scattering$ $rate$: The general
expressions for the optical dipole potential and the scattering rate
are given by \cite{one}
\begin{eqnarray}
U_{dip}\simeq\frac{\hbar\Gamma^{2}}{8\Delta}\frac{I}{I_{s}},\nonumber \\
\gamma_{P}\simeq\frac{\Gamma^{3}}{8\Delta^{2}}\frac{I}{I_{s}},
\end{eqnarray}
where $\Gamma$ is the spontaneous decay rate of the excited level,
$\Delta$ is single photon detuning, $I_{s}=2\pi^{2}\hbar c
\Gamma/3\lambda^{3}$ is the saturation intensity ($\lambda$ is
optical wavelength and $c$ is the speed of light). If the single
photon detuning is larger ($\Delta\gg\nu_{FS}$) than the excited
state fine structure intervals for alkali-metal, the Raman coupling
strength is
\begin{eqnarray}
\Omega\simeq\frac{\Gamma^{2}}{8\Delta}\frac{\nu_{FS}}{\Delta}\frac{\sqrt{I_{1}I_{2}}}{I_{s}},
\end{eqnarray}
where $\nu_{FS}$ is the excited-state fine-structure splitting,
$I_{1}$ and $I_{2}$ are the intensities of two Raman beams. The
dipole potential scales as $I/\Delta$, whereas the scattering rate
scales as $I/\Delta^{2}$. Therefore, optical dipole traps usually
use large detunings and high intensities to keep the scattering rate
as low as possible at a certain potential depth. However, the rate
of the Raman coupling strength and the scattering rate is close to a
constant $\propto\nu_{FS}/\Gamma$ when the single photon detuning is
larger than the excited state fine structure. Therefore it is
useless to increase the detuning $\Delta$ to improve the ratio
between the Raman coupling strength and the scattering rate
\cite{two}. However, the moderate strength of the Raman coupling
still can be obtained during the atomic lifetime due to photon
scattering in the optical dipole trap for the larger single photon
detuning in the paper, since rubidium has 15 nm the excited-state
fine structure spitting.

\end{document}